\documentclass[apj,square,numbers]{emulateapj}
\usepackage{graphicx}
\usepackage{latexsym}
\usepackage{color}
\usepackage[colorlinks, linkcolor=blue, citecolor=blue, urlcolor=blue]{hyperref}
\usepackage{apjfonts}
\usepackage{multirow}

\begin{document}

%\title{Effects on CMB Measurements of Primordial Gravitational Waves from Anisotropies of Polarization Rotation Angle}
\title{Primordial Gravitational Waves Measurements and Anisotropies of CMB Polarization Rotation}

\author{Si-Yu Li$^{1}$}
\author{Jun-Qing Xia$^{2}$}
\author{Mingzhe Li$^{3}$}
\author{Hong Li$^{2}$}
\author{Xinmin Zhang$^{1}$}
\affil{$^1$Theory Division, Institute of High Energy Physics, Chinese Academy of Science, P. O. Box 918-4, Beijing 100049, P. R. China}
\affil{$^2$Key Laboratory of Particle Astrophysics, Institute of High Energy Physics, Chinese Academy of Science, P. O. Box 918-3, Beijing 100049, P. R. China}
\affil{$^3$Interdisciplinary Center for Theoretical Study, University of Science and Technology of China, Hefei, Anhui 230026, P. R. China}

\begin{abstract}

Searching for the signal of primordial gravitational waves in the B-modes (BB) power spectrum is one of the key scientific aims of the cosmic microwave background (CMB) polarization experiments. However, this could be easily contaminated by several foreground issues, such as the interstellar dust grains and the galactic cyclotron electrons. In this paper we study another mechanism, the cosmic birefringence, which can be introduced by a CPT-violating interaction between CMB photons and an external scalar field. Such kind of interaction could give rise to the rotation of the linear polarization state of CMB photons, and consequently induce the CMB BB power spectrum, which could mimic the signal of primordial gravitational waves at large scales. With the recently released polarization data of BICEP2 and the joint analysis data of BICEP2/Keck Array and Planck, we perform a global fitting analysis on constraining the tensor-to-scalar ratio $r$ by considering the polarization rotation angle [$\alpha(\hat{\bf n})$] which can be separated into a background isotropic part [$\bar{\alpha}$] and a small anisotropic part [$\Delta\alpha(\hat{\bf n})$]. Since the data of BICEP2 and Keck Array experiments have already been corrected by using the ``self-calibration'' method, here we mainly focus on the effects from the anisotropies of CMB polarization rotation angle. We find that including $\Delta\alpha(\hat{\bf n})$ in the analysis could slightly weaken the constraints on the tensor-to-scalar ratio $r$, when using current CMB polarization measurements. We also simulate the mock CMB data with the BICEP3-like sensitivity. Very interestingly, we find that if the effects of the anisotropic polarization rotation angle can not be taken into account properly in the analysis, the constraints on $r$ will be dramatically biased. This implies that we need to break the degeneracy between the anisotropies of the CMB polarization rotation angle and the CMB primordial tensor perturbations, in order to measure the signal of primordial gravitational waves accurately.

\end{abstract}
%\pacs{98.80.Cq, 98.80.Es, 11.30.Cp, 11.30.Er}

\maketitle

%Introduction==========================================================

\section{Introduction}

Observations of the Cosmic Microwave Background (CMB) with high precisions are of great importance in modern cosmology, since it is the most powerful way to understand the origin and evolution of our Universe.
It is commonly believed that currently observed CMB anisotropies and large scale structure are originated from the quantum fluctuations at very early times preceding the hot expansion. Theoretical hypothesis about this very early universe includes inflation \citep{guth1981,as1982,linde1982}, ekpyrotic/cyclic universe \citep{khoury2001} and so on. According to these theories the Universe was smoothed and flattened at the early epoch, and at the same time the quantum fluctuations from vacuum extended outside the horizon and formed the primordial perturbations, i.e., the initial conditions of cosmic structure formation.

The primordial perturbations can be decomposed into three types: scalar, vector and tensor perturbations. The scalar perturbations cause the anisotropies of CMB and the structure formation of large scale structure, while the vector perturbations, correspond to vortices decline rapidly during the expanding universe and can be neglected at later time (see e.g. \citet{Mukhanov:2005book}). More interestingly, the tensor perturbations, i.e., the primordial gravitational waves \citep{Starobinsky:1979,Rubakov:1982,Fabbri:1983,Abbott:1984} leave the unique imprint on the large scales power spectrum of CMB B-modes polarization \citep{Kami1997}. Therefore, probes of CMB primordial B-modes polarization are always treated as crucial ways to test the early universe models. Searching for the primordial gravitational waves is one of the key scientific goals of the CMB polarization experiments, such as the BICEP3 \citep{bicep3}, the EBEx \citep{ebex}, the SPT polarization \citep{sptpol} and the ACT polarization \citep{actpol}.

In order to describe the primordial tensor perturbations, we usually use the amplitude parameter, the tensor-to-scalar ratio, $r\equiv{P_T}/{P_S}$, where $P_T$ and $P_S$ denote the power spectra of tensor and scalar perturbations. Due to the precision limitations of current CMB experiments, the constraints on $r$ are still weak and consistent with zero very well \citep{B03,WMAP9,QUIET,bicep2013,Plancktotal,Planckfit2013}. In 2014, the BICEP2 Collaboration published their final three-year CMB observational data and announced the detection of CMB B-modes polarization at scales $20<\ell<340$ \citep{bicep2014}, which corresponds to $r\sim0.2$, and the null detection ($r=0$) is disfavored at about $7\,\sigma$.

However, few months later, the Planck Collaboration measured the angular power spectra of the polarized thermal emission from diffuse Galactic dust, which is the main foreground present in measurements of the CMB polarization at frequencies above 100 GHz \citep{wmapdust2011}, and pointed out that the contribution of thermal dust on the CMB BB power spectrum has been significantly underestimated in the BICEP2 foreground analyses \citep{planckdust2014}. They found that the magnitude of the BB power spectrum, contributed by the polarized thermal dust emission, at 150 GHz in the sky field observed by the BICEP2 experiment is at the same level as the one predicted by the early universe model with the primordial gravitational waves $r\sim0.2$. When the polarized dust contribution is properly taken into account, the joint analysis of the BICEP2/Keck and Planck Collaborations preferred the null detection of the primordial gravitational waves and obtained the upper limit of the tensor-to-scalar ratio $r<0.11$ at 95\% confidence level \citep{BKP2015}.

Besides the polarized thermal dust emission, the cosmic birefringence can also generate the CMB B-modes polarization at large scales. This phenomenon can be introduced by the interaction between CMB photons and the external scalar field $p_{\mu}$ through the Chern-Simons (CS) term $\mathcal{L}_{\rm CS}\sim p_{\mu}A_{\nu}\tilde{F}^{\mu\nu}$. Here $\tilde{F}^{\mu\nu}=(1/2)\epsilon^{\mu\nu\rho\sigma}F_{\rho\sigma}$ is the dual of the electromagnetic tensor. This coupling is gauge invariant if $\partial_{\nu} p_{\mu}=\partial_{\mu} p_{\nu}$. This is possible if $p_{\mu}$ is a constant field over the spacetime or arises from the derivative of a cosmic scalar field $\phi$. The scalar field $\phi$ can be identified as the dark energy in the quintessential baryo-/leptogenesis \citep{Li:2002,Li:2003} and as the Ricci scalar $R$ in the gravitational baryo-/leptogenesis \citep{Li:2004,Davoudiasl:2004gf}.
The CS term violates the Lorentz and CPT symmetries spontaneously if $p_{\mu}$ has nonzero background values.
This induces rotations of polarization directions of the propagating photons \citep{Carroll1990,Carroll1998} and productions of  CMB TB and EB cross-correlations, even though they are absent before the recombination in the traditional CMB theory \citep{Lue:1999,Feng2005,Feng2006,Li2007,Xia2008a}. The crucial point for this paper is that this phenomenon provides a mechanism to produce the CMB B-modes,  alternative to the primordial gravitational waves and weak lensing. Because the rotation converts part of E-modes polarization to B-modes polarization, even the primordial B-modes polarization is absent, sizable CMB B-modes power spectrum can still be obtained from the E-modes power spectrum through the rotation. As shown in \citet{Xia2010}, the cosmological birefringence degenerates with primordial tensor mode perturbations at large scales and with the gravitational lensing at small scales, respectively. And due to this degeneracy, considering the anisotropic polarization rotation could slightly lower the best-fit value of $r$ and relax the tension on the constraints of $r$ between from BICEP2 and from Planck 2013 data \citep{Li2015}, even without the help of the isotropic part of the rotation angle,  which is removed by the ``self-calibration'' method in the BICEP2 data analysis \citep{bicep2014}.

Therefore, it is important and necessary to study the influences of the polarization rotation angle on the measurements of CMB primordial tensor perturbations using the current and future CMB experiments. Recently, \citet{XiaSys} found that constraints on the tensor-to-scalar ratio could be significantly biased, when the isotropic rotation angle can not be taken into account properly. In this paper, our main task is to focus on the influences of the anisotropic polarization rotation angle on the detection of primordial gravitational waves by performing the global fitting with the latest CMB datasets and simulated mock data of future CMB experiments included. This paper is organized as follows: in section \ref{theory} we briefly introduce the cosmic birefringence mechanism and the effects on CMB polarization power spectra from the isotropic and anisotropic polarization rotation angle. Section \ref{result} contains our main results from the current observations and future measurements, while section \ref{summary} is dedicated to the conclusions and discussions.

%CPT Theory===========================================================

\section{CMB Polarization Rotation Angle}\label{theory}

In this section we briefly introduce the cosmic birefringence which has been widely studied in the literatures \citep{Carroll1990,Carroll1998,Lue:1999,Feng2005,Feng2006,Li2007,Xia2008a,Li:2008,Kamionkowski:2008,Yadav:2009eb,Caldwell:2011}. We consider the CS coupling between the cosmic scalar field $\phi$ and the electromagnetic field:
\begin{equation}\label{coupling}
\mathcal{L}_{\rm CS}= \partial_{\mu}f(\phi)A_{\nu}\tilde{F}^{\mu\nu}~,
\end{equation}
where $f$ is an arbitrary function of $\phi$. As mentioned before, the CS term violates Lorentz and CPT symmetries but not the gauge symmetry of the electromagnetic field, so will not change the number of dynamical components of photons. However, the states with opposite helicity propagates with different velocities, so that the polarization vectors of photons will not be parallel transported along the light rays. Consequently the observed polarization directions are changed by rotations when compared with the original directions at the source.

In terms of the Stokes parameters $Q$ and $U$, the rotated linear polarization state (denoted by a prime) is related to the unrotated one through
\begin{equation}\label{rotate}
(Q'\pm iU')={\rm exp}(\pm i2\alpha)(Q\pm iU)~.\label{rotation1}
\end{equation}
The rotation angle $\alpha$ corresponds to the CPT violation and is frequency independent. It can be calculated by the integration along the light rays \citep{Li:2008}
\begin{equation}\label{ro}
\alpha=\int^s_o\partial_{\mu}f(\phi) dx^{\mu}(\lambda)=f(\phi_s)-f(\phi_0)~,
\end{equation}
where $\lambda$ is the affine parameter of the light ray, the subscripts $s$ and $o$ represent the source and the observer, respectively. Naturally, we observe CMB photons at a single point, hence $\phi_o$ should be a constant. However the dynamical field $\phi$ is expected to fluctuate on the last scattering surface (LSS). The observer at the fixed position receives CMB photons from all directions and will find that the rotation angle varies across the sky, as pointed out in \citet{Li:2008}. The anisotropies are proportional to the distribution of $\phi$ on the LSS, which means we can separate the rotation angle $\alpha$ into the isotropic part and the small anisotropic one as follows
\begin{equation}
\alpha(\hat{\bf n}) = \bar{\alpha} + \Delta\alpha(\hat{\bf n})~,
\end{equation}
with
\begin{eqnarray}\label{dphi}
\bar{\alpha}&=&f[\bar{\phi}(\eta_{\rm LSS})]-f(\phi_0)~,\nonumber\\
\Delta\alpha(\hat{\bf n})&=&\frac{df}{d\phi}\delta\phi(\vec{x}_{\rm LSS}, \eta_{\rm LSS})~,
\end{eqnarray}
where $\eta_{\rm LSS}$ and $\vec{x}_{\rm LSS}$ denote the position and the conformal time on the last scattering surface. For the isotropic part $\bar{\alpha}$, the first evidence in terms of the full CMB datasets was done in \citet{Feng2006} and stimulated many interests in this field (see \citet{Li2007,Xia2008a,WMAP5,Xia2008b,Wu:2009,Brown:2009,WMAP7,Xia2010,Liu:2006,XiaPlanck,XiaSys,WMAP9,
Geng2007,Cabella2007,Kostelecky2007,Kahniashvili2008,Finelli2009,Li2009,gruppuso2013,zhao2013,zhao2015}, and references within).

Assuming the anisotropic rotation angle $\Delta\alpha(\hat{\bf n})$ is a zero mean field defined on two dimensional sphere, it can be decomposed in terms of spherical harmonics:
\begin{eqnarray}\label{sphericalharmonics}
\Delta\alpha(\hat{\bf n})&=&\sum b_{\ell m}Y_{\ell m}(\hat{\bf n})~.
\end{eqnarray}
Furthermore, if we assume the perturbation field $\Delta\alpha(\hat{\bf n})$ is a Gaussian random field and satisfies the statistical isotropy, so we can define the angular power spectrum of the anisotropic rotation angle:
\begin{equation}
\langle b_{\ell m}b^{*}_{\ell' m'}\rangle=C^{\alpha\alpha}_\ell \delta_{\ell\ell'}\delta_{mm'}~.
\end{equation}
In this paper we set $f(\phi)={c\phi}/{M}$  for simplicity, in which $M$ represents the cut-off mass in the effective field theory and $c$ stands for the dimensionless coupling coefficient. We find the relationship between the angular power spectrum of anisotropic rotation angle $C_\ell^{\alpha\alpha}$ and the power spectrum of the scalar field $P_{\phi}$ at the LSS
\begin{eqnarray}\label{connect1}
\langle \phi_{\vec{k}} \phi^{*}_{\vec{k'}}\rangle\bigg{|}_{\rm LSS} &=& \frac{2\pi^2}{k^3}P_{\phi}(k)\delta^3(\vec{k}-\vec{k'})~,\\
C^{\alpha\alpha}_{\ell} &=& \frac{4\pi c^2}{M^2}\int\frac{dk}{k}P_{\phi}(k)j^2_\ell(k\eta_0-k\eta_{s})
\end{eqnarray}
where $\phi_{\vec{k}}$ is the Fourier transform of $\delta\phi$ and $j_\ell(x)$ is the spherical Bessel function.

\citet{ZhaoLi2014} has considered several specific dark energy models, such as the massless scalar field and the quintessence field, and then calculated their power spectra explicitly. In this paper we study the constraint on the perturbations of this scalar field through the polarization rotation angle in a model-independent way. It is convenient to expand its power spectrum on the super hubble scale, just as what we usually do in the study of slow-roll inflation,
\begin{equation}\label{connect2}
{P}_{\phi} \equiv {P}_{\rm CPT}=A_{\rm CPT}\left(\frac{k}{k_0}\right)^{n_{\rm CPT}}~,
\end{equation}
where ${A}_{\rm CPT}$ , $n_{\rm CPT}$ and $k_0$ represent the amplitude and spectral index of the power spectrum and the pivot scale, respectively. Due to the precision limitation of current CMB experiments, we set $n_{\rm CPT}$ to be zero and neglect the higher order running parameters of the spectral index [$d\ln n_{\rm CPT}/d\ln k$] in the following calculations.

In order to investigate the influence on the CMB polarization power spectra due to the cosmic polarization rotation angle, we firstly define two-point correlation function using the following relation
\begin{equation}
C^{\alpha\alpha}(\beta)\equiv\langle\delta\alpha(\hat{\bf n})\delta\alpha(\hat{\bf n'})\rangle=\sum_{\ell}\frac{2\ell+1}{4\pi}C_\ell^{\alpha\alpha} P_\ell(\cos\beta)~,
\end{equation}
where $\beta$ corresponds to the angle between these two directions, $\cos(\beta)=\hat{\bf n}\cdot\hat{\bf n}'$. Combining Eqs. (\ref{rotate}) and (\ref{sphericalharmonics}), one can express the rotated CMB polarization power spectra $C'_{\ell}$ in terms of the unrotated ones $C_{\ell}$ (see \citet{Li:2013} for details):
\begin{widetext}
\begin{eqnarray}\label{newCMB}
&&{C'_{\ell}}^{EE}+{C'_{\ell}}^{BB}=\exp{[-4C^{\alpha\alpha}(0)]}\sum_{\ell'}\frac{2\ell'+1}{2}(C_{\ell'}^{EE}+C_{\ell'}^{BB})
\int^{1}_{-1}d^{\ell'}_{22}(\beta)d^{\ell}_{22}(\beta)e^{4C^{\alpha\alpha}(\beta)}d\cos(\beta)\nonumber\\
&&{C'_{\ell}}^{EE}-{C'_{\ell}}^{BB}=\cos(4\bar{\alpha})\exp{[-4C^{\alpha\alpha}(0)]}\sum_{\ell'}\frac{2\ell'+1}{2}(C_{\ell'}^{EE}
-C_{\ell'}^{BB})\int^{1}_{-1}d^{\ell'}_{-22}(\beta)d^{\ell}_{-22}(\beta)e^{-4C^{\alpha\alpha}(\beta)}d\cos(\beta)\nonumber\\
&&{C'_{\ell}}^{EB}=\sin(4\bar{\alpha})\exp{[-4C^{\alpha\alpha}(0)]}\sum_{\ell'}\frac{2\ell'+1}{4}(C_{\ell'}^{EE}-C_{\ell'}^{BB})
\int^{1}_{-1}d^{\ell'}_{-22}(\beta)d^{\ell}_{-22}(\beta)e^{-4C^{\alpha\alpha}(\beta)}d\cos(\beta)\nonumber\\
&&{C'_{\ell}}^{TE}=\cos(2\bar{\alpha})\exp{[-2C^{\alpha\alpha}(0)]}\sum_{\ell'}\frac{2\ell'+1}{2}C_{\ell'}^{TE}\int^{1}_{-1}
d^{\ell'}_{02}(\beta)d^{\ell}_{20}(\beta)d\cos(\beta)\nonumber=C_{\ell}^{TE}\cos(2\bar{\alpha})e^{-2C^{\alpha\alpha}(0)}\nonumber\\
&&{C'_{\ell}}^{TB}=\sin(2\bar{\alpha})\exp{[-2C^{\alpha\alpha}(0)]}\sum_{\ell'}\frac{2\ell'+1}{2}C_{\ell'}^{TE}\int^{1}_{-1}
d^{\ell'}_{02}(\beta)d^{\ell}_{20}(\beta)d\cos(\beta)=C_{\ell}^{TE}\sin(2\bar{\alpha})e^{-2C^{\alpha\alpha}(0)}~,
\end{eqnarray}
\end{widetext}
where $C^{\alpha\alpha}(0)\equiv\sum_\ell(2\ell+1)C_\ell^{\alpha\alpha}/4\pi$ is the auto-correlation of the anisotropic polarization rotation angle. Note that, in the calculations we have assumed the original CMB fields and anisotropic rotation angle field are statistically isotropic Gaussian random fields.

\begin{figure}[t]
\begin{center}
  \includegraphics[scale=0.62]{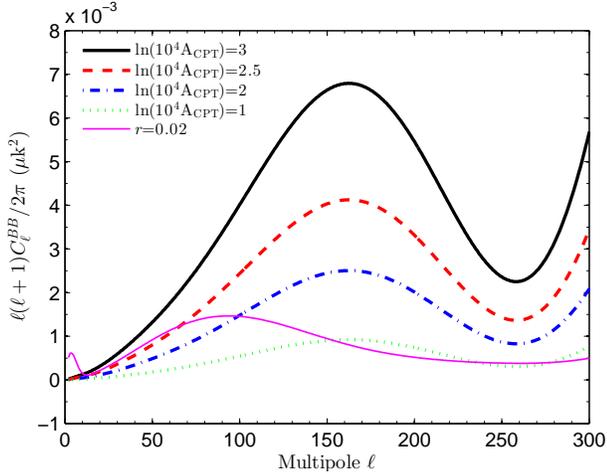}
  \caption{The theoretical CMB BB power spectra for different values of $\ln[10^4A_{\rm CPT}]$. We also show the theoretical CMB BB power spectrum with the tensor-to-scalar ratio $r=0.02$ (magenta thin line) for comparison.}\label{plotbb}
\end{center}
\end{figure}

In the Eqs. (\ref{newCMB}) we have separated the influences of isotropic and anisotropic polarization rotation angle. It is apparently that they both get involved in all scales of CMB polarization power spectra. The isotropic rotation angle copies the original CMB EE power spectrum, zooms it out and then prints it onto the rotated BB power spectrum. It will inevitably create additional CMB TB and EB power spectra which should be vanished in normal CMB theories. We can see the terms $\cos(\bar{\alpha})$ or $\sin(\bar{\alpha})$ appear as coefficients in almost all equations. This suggests us that the isotropic rotation angle should perform a global modulation on rotated CMB power spectra if it exists. However, the situation for the anisotropic one is somehow different. Non-zero TB and EB spectra are not longer necessary if the isotropic part [$\bar\alpha$] vanishes, which is consistent with some current CMB experiments. The BICEP2/Keck Array collaboration has used the ``self-calibration'' method for the detector polarization orientations, and as a result, any signal of the isotropic polarization rotation angle has been eliminated from data. Fortunately, the anisotropic part is not affected by this calibration method, and will still contaminate the observations of the primordial gravitational waves. In figure \ref{plotbb}, we plot the CMB BB power spectra for different values of $\ln[10^4A_{\rm CPT}]$ with $\bar{\alpha}=0$. For comparison, we also show the expected CMB BB power spectrum generated by the primordial tensor perturbations $r=0.02$. Interestingly, we find that, even the isotropic rotation angle is zero, the anisotropic polarization rotation angle with $\ln[10^4A_{\rm CPT}] \sim 2.5$ could still generate the CMB BB power spectrum to mimic CMB primordial B-modes signal with $r=0.02$ at large scales. At scales $\ell < 20$, the contribution on BB power spectrum from the anisotropic polarization rotational angle can be negligible. Therefore, probing the reionization bump of CMB BB power spectrum at very large scale may provide a smoking gun for distinguishing the polarization rotational angle and primordial gravitational waves.

%Method&Data===========================================================

\section{Numerical Results}\label{result}

In this paper we perform a global analysis to all the power spectra of the CMB data with the public available Markov Chain Monte Carlo package {\tt CosmoMC} \citep{Lewis:2002}, which has been modified to compute the rotated CMB polarization power spectra discussed above. We assume the purely adiabatic initial conditions and impose the flatness condition motivated by Inflation. Our basic parameter space is: ${\bf P} \equiv (\omega_{b}, \omega_{c}, \Omega_\Lambda, \tau, n_{s}, A_{s}, r)$, where $\omega_{b}\equiv\Omega_{b}h^{2}$ and $\omega_{c}\equiv\Omega_{c}h^{2}$ are the physical baryon and cold dark matter densities relative to the critical density, $\Omega_\Lambda$ is the dark energy density relative to the critical density, $\tau$ is the optical depth to re-ionization, $A_{s}$ and $n_{s}$ characterize the primordial scalar power spectrum, $r$ is the tensor to scalar ratio of the primordial spectrum. For the pivot of the primordial spectrum we set $k_{\rm s0}=0.05\,$Mpc$^{-1}$. For the polarization rotation angle, we set the isotropic rotation angle $\bar\alpha\equiv0$ and then have one free parameter, $A_{\rm{CPT}}$, which represent the amplitude of power spectrum of the anisotropic rotation angle. Furthermore, in our analysis we include the CMB lensing effect, which also produces B-modes from E-modes \citep{Z:lensing}, when we calculate the theoretical CMB power spectra.

\subsection{Current CMB Observations}

In our calculations we mainly use the polarization power spectra from BICEP2 experiments and the joint analysis data of BICEP2/Keck Array and Planck experiments:
\begin{itemize}

\item The {\it Background Imaging of Cosmic Extragalactic Polarization} (BICEP2) \citep{bicep2014}, locating at the South Pole, specializes in searching polarized CMB signal at large scales with low angular resolution, concentrating 150 GHz on a roughly 1$\%$ patch of sky at high Galactic latitudes. Recently the BICEP2 collaboration released their three-year data accumulated from 2010 to 2012 and announced the detection of CMB B-modes polarization at scales $20<\ell<340$. It is noteworthy that they claimed that the CMB EB power spectrum is only used for the ``self-calibration'' of the detector polarization orientations \citep{keating}. Any polarization rotation has been removed from the results.

\item The products of the Joint Analysis of BICEP2/Keck Array and Planck Data (BKP2015) \citep{BKP2015}. Very recently, BICEP2/Keck Array and Planck Collaborations have re-analysed their data and published the CMB BB power spectrum at scales $20<\ell<340$, in order to investigate the influence on the CMB B-modes measurements from the polarized thermal dust emission. They convert the Planck maps into the format usable for BICEP2/Keck Array experiments, and make cross-spectra between the combined BICEP2/Keck maps and Planck maps of the all polarized bands. Similar with the BICEP2 data, the ``self-calibration'' method was also applied in the data analysis.

\end{itemize}

%Current Result========================================================

\begin{figure}[t]
\begin{center}
  \includegraphics[scale=0.56]{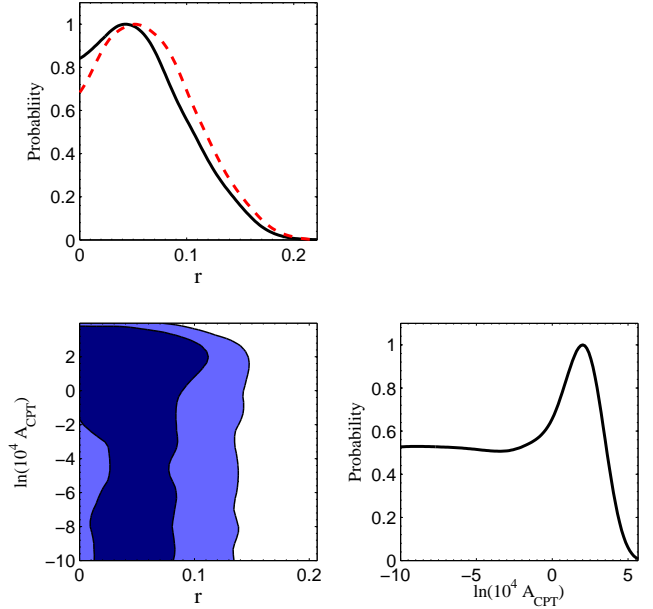}
  \caption{One- and two-dimensional constraints on the tensor-to-scalar ratio and the anisotropic polarization rotation angle from the BICEP2 data (black solid lines). For comparison, we also show the constraint on $r$ from the BICEP2 data with $A_{\rm CPT}=0$ (red dashed line).}\label{r_Acpt_tri}
\end{center}
\end{figure}

Due to the ``self-calibration'' method used in the BICEP2/Keck Array experiments, the constraints on the isotropic rotation angle [$\bar\alpha$] are quite consistent with zero \citep{Li2015}. Therefore, we only use these data to study the anisotropies of the polarization rotation angle. Furthermore, in order to take the contribution of polarized thermal dust emission on the CMB B-modes power spectrum into account properly, we add one more nuisance parameters $A_{\rm{dust}}$ in our calculations and consider the polarized dust BB power spectrum satisfies $C^{\rm{BB}}_{\ell,\rm{dust}}\propto A_{\rm{dust}}\ell^{-2.42}$ \citep{planckdust2014}.

In figure \ref{r_Acpt_tri} we show the constraints on the tensor-to-scalar ratio and the anisotropic polarization rotation angle, and their 2-dimensional contours from the BICEP2 data. Firstly, we neglect the contribution on the CMB B-modes power spectrum from the polarization rotation angle, which means $A_{\rm CPT}\equiv0$. Due to the strong degeneracy between $r$ and $A_{\rm dust}$, the huge uncertainty of the polarized dust B-modes power spectrum significantly reduces the constraining power of BICEP2 data on the tensor-to-scalar ratio $r$. The obtained $2\sigma$ upper limit is $r < 0.12$ and there is a peak around $r\sim0.05$, shown as the red dashed line in figure \ref{r_Acpt_tri}. There is no signal for the primordial gravitational waves, which is consistent with the results from the BKP2015 data \citep{BKP2015}.

\begin{figure}[t]
\begin{center}
  \includegraphics[scale=0.66]{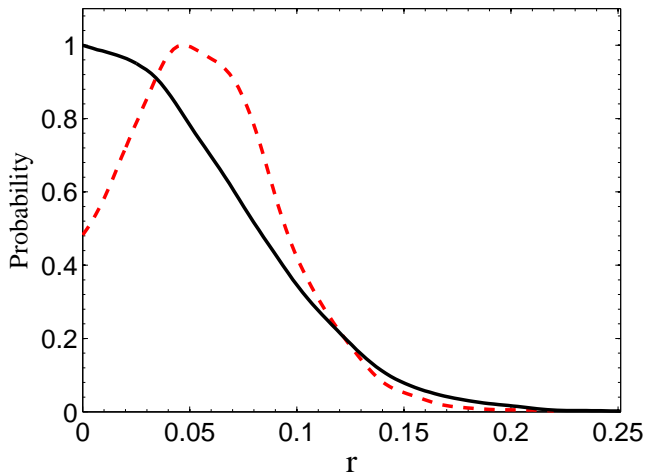}
  \caption{One-dimensional constraints on the tensor-to-scalar ratio from the BKP2015 data with (black solid line) and without (red dashed line) the effect of the anisotropic polarization rotation angle included.}\label{r_BKP2015}
\end{center}
\end{figure}

When we include the effect of anisotropic polarization rotation angle in the calculations, the constraint on the tensor-to-scalar ratio is slightly different. As we mentioned in \citet{Li2015}, the anisotropic polarization rotation angle would also contribute to the CMB BB power spectrum and partly explain the CMB B-modes data of BICEP2. In this case, the 1-dimensional distribution of $r$ slightly moves towards to the smaller value, due to this degeneracy, see the black solid line in figure \ref{r_Acpt_tri}. And the minimal value of $\chi^2$ becomes slightly lower, $\Delta\chi^2 \sim {-1}$. We also check this result using the latest BKP2015 data, which is shown in figure \ref{r_BKP2015}. When we include the contribution on CMB B-modes power spectrum from the anisotropic polarization rotation angle, the upper limit of tensor-to-scalar ratio is $r < 0.13$ (95\% C.L.) and the peak structure totally disappears. These results imply that even considering the effect of polarized thermal dust emission, the anisotropic polarization rotation angle can still affect the constraint on the tensor-to-scalar ratio slightly.

In figure \ref{r_Acpt_tri} we also show the one-dimensional constraint on the amplitude parameter $A_{\rm CPT}$. The current BICEP2 data still do not have enough constraining power to limit $A_{\rm CPT}$ accurately, so the obtained constraint on $A_{\rm CPT}$ is consistent with zero, namely the 95\% upper limit is $\ln[10^4A_{\rm CPT}] < 3.8$. However, we interestingly find that there is a peak around $\ln[10^4A_{\rm CPT}] \sim 2.5$. As we show in figure \ref{plotbb}, the model with $\ln[10^4A_{\rm CPT}] \sim 2.5$ could mimic the contribution of CMB B-modes power spectrum of primordial gravitational waves with $r=0.02$ at large scales. And this degeneracy moves the distribution peak of $r$ towards to the lower value, which is similar with the results in \citet{Li2015}.

Finally, we plot the two-dimensional contour on the panel ($r$,$\ln[10^4A_{\rm CPT}]$) in figure \ref{r_Acpt_tri}. Apparently, these two parameters are slightly correlated. The model with a larger value of $\ln[10^4A_{\rm CPT}]$ provides more power of CMB B-modes, and consequently the contribution from the primordial gravitational waves is suppressed, namely the model with $r=0$ is more favored by the BICEP2 data. Conversely, if we limit the parameter $\ln[10^4A_{\rm CPT}]$ to be a very small value, the large value of $r$ will be needed, in order to match the BICEP2 data.

\begin{figure}[t]
\begin{center}
  \includegraphics[scale=0.66]{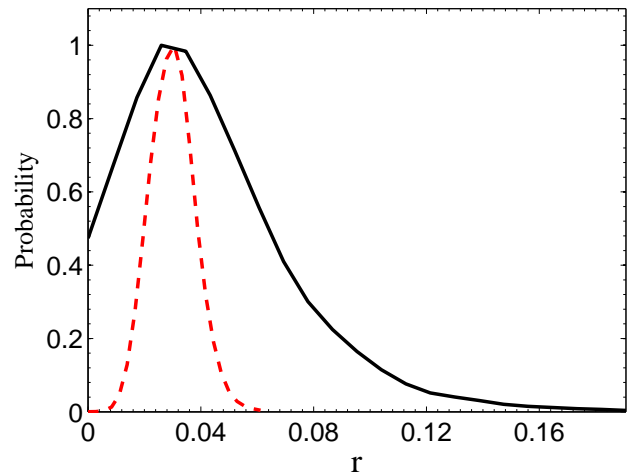}
  \caption{One-dimensional constraints on the tensor-to-scalar ratio from the simulated mock CMB data with (black solid line) and without (red dashed line) the effect of the anisotropic polarization rotation angle included.}\label{simulate1}
\end{center}
\end{figure}

\subsection{Future CMB Measurements}

Since the current CMB data are not accurate enough, the degeneracy between $r$ and $\ln[10^4A_{\rm CPT}]$ is not very clearly. But this could be important in the future CMB B-modes experiments. Therefore, it is worthwhile discussing the effect of non-zero anistropic polarization rotation angle on the measurement of CMB primordial B-modes signal for the future experiments. Following \citet{XiaPlanck}, we simulate the future mock CMB power spectra with BICEP3-like sensitivity: $f_{\rm sky}=0.05$ sky converge, the isotropic noise with variance $\Delta_{\rm T} = \Delta_{\rm P} / 2 = 0.13 \mu K^2$, a symmetric Gaussian beam of $25$ arcminutes full-width half-maximum (FWHM) \citep{bicep3}, and the maximum multipoles $\ell_{\rm max} = 1000$. Here, we neglect the systematic errors of future CMB measurements and the contaminations from the CMB various foregrounds, such as the thermal dust emission and the synchrotron emission. The basic fiducial model we choose is the best-fit Planck model \citep{Planckfit2015}: $\Omega_{b}h^2=0.02225$, $\Omega_{c}h^2=0.1194$, $100\Theta_s=1.04094$, $\tau=0.079$, $n_{s}=0.9682$, $\log{[10^{10}A_{s}]}=3.08919$ at $k_{\rm s0}=0.002\,$Mpc$^{-1}$.

Firstly, we simulate the mock data using the basic fiducial model and the tensor-to-scalar ratio $r=0.03$. When using the basic seven cosmological parameters ($A_{\rm CPT}\equiv0$) to fit the mock data, we obtain a $3\sigma$ detection of non-zero value of tensor-to-scalar ratio: $r=0.03 \pm 0.01$ (68\% C.L.). However, when we set the parameter $\ln[10^4A_{\rm CPT}]$ to be free, the constraint on $r$ is obviously weakened, $r=0.03 \pm 0.03$ (68\% C.L.) and $r<0.1$ (95\% C.L.), as shown in figure \ref{simulate1}. Note that, the standard deviations of $r$ we obtain are apparently underestimated, since in the calculations we do not consider any contamination from the CMB foregrounds. This result implies that in the future CMB experiments, if we want to measure the signal of CMB primordial B-modes accurately, we need to know very well about the anisotropic polarization rotation angle. Otherwise, the degeneracy among them will contaminate the signal of primordial gravitational waves significantly.

\begin{figure}[t]
\begin{center}
  \includegraphics[scale=0.66]{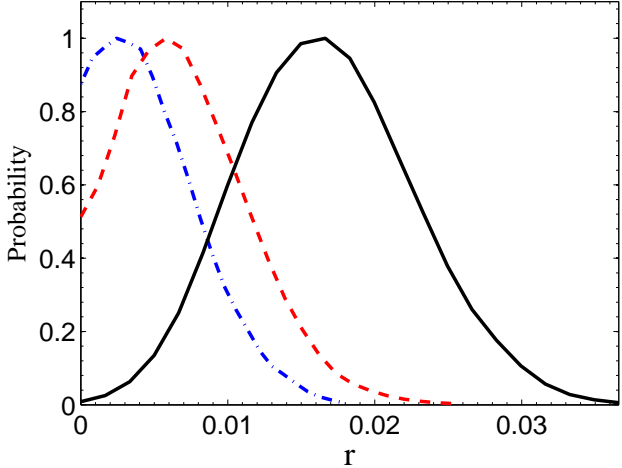}
  \caption{The one-dimensional posterior distributions of the tensor-to-scalar ratio $r$ derived from the simulated mock CMB data without including the anisotropic rotation angle. The anisotropic rotation angles considered in the fiducial model are: $\ln[10^4A_{\rm CPT}]=0$ (blue dash-dotted line), $\ln[10^4A_{\rm CPT}]=1$ (red dashed line), and $\ln[10^4A_{\rm CPT}]=2$ (black solid line).}\label{simulate2}
\end{center}
\end{figure}

We also simulate another mock data using the basic fiducial model and the non-zero anisotropic polarization rotation angle. Here, we consider three fiducial values $\ln[10^4A_{\rm CPT}]=0,1,2$. When we include the parameter of anisotropic polarization rotation angle in the calculations, the fiducial values of input parameters are always recovered. However, if we neglect the effect of anisotropic rotation angle at all $A_{\rm CPT}\equiv0$, the contribution of non-zero anisotropic rotation angle on the CMB BB power spectrum will be wrongly considered as the signal of the primordial CMB tensor perturbations. Since the fiducial model $r=0$, the BB power spectrum due to the primordial tensor B-modes should vanish. But in the presence of non-zero fiducial rotation angle ($\ln[10^4A_{\rm CPT}]=0,1,2$), the BB power spectrum should be non-vanishing. If we force the anistropic rotation angle to be zero in the analysis, the value of $r$ will be enlarged to match the mock non-zero BB power spectrum. In figure \ref{simulate2}, we find that the larger the fiducial value of $\ln[10^4A_{\rm CPT}]$ is, the larger the obtained central value of $r$ becomes. For example, in the calculation with the fiducial model $\ln[10^4A_{\rm CPT}] = 2$, if we fix $A_{\rm CPT}=0$, the obtained constraint on the primordial tensor perturbations $r=0.017 \pm 0.006$ (68\% C.L.). The significance of this fake signal is about $3\sigma$ confidence level. Therefore, if we do not take the effect of polarization rotation angle into account properly, the constraints on the tensor-to-scalar ratio $r$ from the future CMB data with the high precision will be significantly biased.

%Summary===============================================================

\section{Conclusions and Discussions}\label{summary}

One of the most important scientific aims in the CMB cosmology is searching for the primordial gravitational waves on the CMB B-modes power spectrum. However, the signal is very small and could be contaminated by other sources easily, such as the CMB foregrounds, polarized thermal dust emission and synchrotron emission. In this paper, we discuss another mechanism, the cosmic birefringence, which can be introduced by the CPT-violating interaction between CMB photons and the external scalar field. This mechanism can also produce the CMB B-modes power spectrum and mimic the signal of primordial gravitational waves at large scales. Here we summarize our main conclusions in more detail:
\begin{itemize}
\item We use the BICEP2 data and the joint analysis data of BICEP2/Keck Array and Planck experiments to constrain the tensor-to-scalar ratio. When we include the contribution of polarized thermal dust emission on the CMB B-modes power spectrum, the constraints on $r$ are consistent with zero at $95\%$ confidence level. Furthermore, when we consider the effect of the anisotropic polarization rotation angle into the analyses, the degeneracy between $r$ and $A_{\rm CPT}$ slightly changes the constraint on $r$. The one-dimensional distribution of $r$ slightly moves towards to the smaller value, since the anisotropic polarization rotation angle would also contribute to the CMB BB power spectrum and partly explain the CMB B-modes data.

\item The current BICEP2 data still do not have enough constraining power to study the polarization anisotropic rotation angle accurately, so the obtained constraint on $A_{\rm CPT}$ is consistent with zero. However, there is a clear peak around $\ln[10^4A_{\rm CPT}]\sim2.5$, which could mimic the contribution of CMB B-modes power spectrum of primordial gravitational waves at large scales, as shown in figure \ref{plotbb}. We also find the slight correlation between $r$ and $\ln[10^4A_{\rm CPT}]$ in the two-dimensional contours.

\item We simulate the mock CMB data using the BICEP3-like sensitivity and study the degeneracy between the tensor-to-scalar ratio and the anisotropic rotation angle. When we neglect the effect of nonzero $A_{\rm CPT}$, the mock data can constrain $r$ tightly. However, when including this effect into analysis, the obtained standard deviation of $r$ is significantly enlarged by a factor of three. This degeneracy will contaminate the signal of primordial gravitational waves obviously.

\item Finally, we simulate the mock data with the different fiducial values. If we neglect the effect of anisotropic rotation angle in the calculations, the contribution of non-zero anisotropic rotation angle on the CMB BB power spectrum will be wrongly considered as the signal of the primordial tensor perturbations. Therefore, we need to take the effect of polarization rotation angle into account properly. Otherwise, the constraints on the tensor-to-scalar ratio $r$ could be biased from the future CMB experiments with high precisions.

\end{itemize}

%Acknowledgments=======================================================

\section*{Acknowledgements}

S. L. and X. Z. are supported in part by the National Science Foundation of China under Grants No. 11121092, No.11375202 and No. 11033005. J.-Q. X. is supported by the National Youth Thousand Talents Program and the NSFC under Grant No. 11422323. M. L. is supported in part by NSFC under Grant No. 11422543, the Fundamental Research Funds for the Central Universities, and the Program for New Century Excellent Talents in University. H. L. is supported in part by the NSFC under Grant No. 11033005 and the youth innovation promotion association project and the Outstanding young scientists project of the Chinese Academy of Sciences. The research is also supported by the Strategic Priority Research Program ``The Emergence of Cosmological Structures'' of the Chinese Academy of Sciences, Grant No. XDB09000000.

%References============================================================
\bibliographystyle{plainnat}

\end{document}